\documentclass[12pt]{article}
\usepackage{epsfig}
\newcommand\bea{\begin{eqnarray}}
\newcommand\eea{\end{eqnarray}}
\begin{document}
\thispagestyle{empty}
\bibliographystyle{unsrt}
\setlength{\baselineskip}{18pt}
\parindent 24pt
\vspace{50pt}

\begin{center}{
{\Large {{\bf Random matrix analysis of network Laplacians} }}
\vskip 1truecm
Sarika Jalan ${^{(a)}}$ and Jayendra N. Bandyopadhyay ${^{(b)}}$

{Max-Planck-Institut f\"{u}r Physik Komplexer Systeme,
N\"{o}thnitzerstr. 38, D-01187 Dresden, Germany}
}
\end{center}

\vskip 1truecm
\begin{abstract}
We analyze eigenvalues fluctuations of the Laplacian of various networks under the 
random matrix theory framework. Analyses of random networks, scale-free networks 
and small-world networks show that nearest neighbor spacing distribution of 
the Laplacian of these networks follow Gaussian orthogonal ensemble  
statistics of random matrix theory. Furthermore, we study nearest neighbor spacing
distribution as a function of the random connections and find that transition to the 
Gaussian orthogonal ensemble statistics occurs at the small-world transition.
\end{abstract}



e-mail address: (a) sarika@pks.mpg.de, (b) jayendra@pks.mpg.de

\section{Introduction}
\label{intro}

In order to understand complex world around us network theory has been getting fast 
recognition. The main concept here is to define complex systems in terms of 
networks of many interacting units. Few examples of such systems are interacting 
molecules in living cell, nerve cells in brain, computers in Internet 
communication, social networks of interacting people, airport networks with flight 
connections, etc \cite{rev-Strogatz,rev-network,rev-Boccaletti,rev-Costa}. To 
understand these networks simple models are introduced; these model networks are 
based on some simple principles, and capture essential features of real systems. 
Mathematically networks are investigated under the framework of graph theory. In 
the graph theoretical terminology, units are called nodes and interactions are 
called edges \cite{graph}.

Random graph model of Erd\"os and R\'enyi (ER) assumes that interaction between the 
nodes are random \cite{erdos}. Recently, with the availability of large maps of 
real world networks, it has been observed that the random graph model is not 
appropriate for studying real world networks. Hence many new models have been 
introduced. Barab\'asi-Albert's scale-free (SF) model \cite{BA} and 
Watts-Strogatz's small-world (SW) model \cite{SW} are the most recognized ones, and 
have contributed immensely in understanding the evolution and behavior of the real 
systems having network structures. SF model captures the degree distribution 
behavior of real world networks and SW model deals with the clustering coefficient 
and diameter. Following these two new models came an outbreak in the field of 
networks. These works have focused on the following aspects: (1) direct studies of 
the real-world networks and measuring their various structural properties such as 
degree distribution, diameter, clustering coefficient, etc., (2) proposing new 
random graph models motivated by these studies, and (3) computer simulations of the 
new models and measuring their properties \cite{rev-network}.

Barab\'asi et.al. investigations of the various real world systems show that they 
are scale-free, which means that the degree distribution $p(k)$, fraction of nodes 
that have $k$ number of connections with other nodes, decays as power law, i.e. 
$p(k) \propto k^{-\gamma}$, where $\gamma$ depends on the topology of the networks. 
The scale-free nature of networks suggests that there exist few nodes with very 
high degree. Some other analysis, by Newman and others, of real-world networks show 
that complex networks have community or module structures \cite{Newman,Amaral}. 
According to these studies, there exists few nodes with very high betweenness which 
are responsible to connect the different communities. This direction of looking at 
the networks focuses on the importance of nodes based on its position in the 
network. On the other hand, ER and SW models emphasize on the random connections in 
the networks; in ER model any two nodes are connected with probability $p$. One of 
the most interesting characteristics of ER model was sudden emergence of various 
global properties; most important one being emergence of giant cluster. For a $p > 
p_c$, while number of nodes in the graph remain constant, giant cluster emerges 
through a phase transition. Further, Watts-Strogatz model shows the small-world 
transition with fine tuning of the number of random connections.

Apart from the above mentioned investigations which focus on direct measurements of 
structural properties of networks, there exists a vast literature demonstrating 
that properties of networks or graphs could be well characterized by the spectrum 
of associated adjacency ($A$) and Laplacian ($L$) matrix \cite{spectrum}. For an 
unweighted graph, adjacency matrix is defined in following way : $A_{ij} = 1$, if 
$i$ and $j$ nodes are connected and {\it zero} otherwise. Laplacian of graph has 
been defined in the various ways (depending upon the normalization) in the 
literature. We follow definition used in \cite{Lap};
\begin{equation} L_{ij} = 1 
~\mbox{for}~ i = j ~\mbox{and}~ L_{ij} = -A_{ij}/\sqrt{d_i d_j}, 
\label{lap} 
\end{equation} 
where $d_i$ denotes the degree of node $i$. For undirected networks, adjacency and 
Laplacian both are symmetric matrices and consequently have real eigenvalues 
\cite{note}. Eigenvalues of graph are called graph spectra and they give 
informations about some basic topological properties of the underlying network. 
During last twenty years several important applications of the 
spectral graph theory in physics and chemistry problems have been discovered
\cite{spectrum,handbook}. For 
example liquid flowing through a system of communicating pipes are described by a 
system of linear differential equations. The corresponding matrix appears to be the 
Laplacian of the underlying graph. Speed of convergence of the liquid flowing 
process towards an equilibrium state is measured by the second smallest eigenvalue 
of $L$ \cite{handbook}. Second smallest eigenvalue of $L$ is also called 
the algebraic connectivity of a graph and is used to understand behavior of 
dynamical processes on the underlying networks \cite{syn-MSF}. Particularly, 
Laplacian spectra of networks have been investigated enormously to understand 
synchronization of coupled dynamics on networks 
\cite{syn-MSF,syn-Atay,syn-spectra}, for example recently extremal eigenvalues of 
the Laplacian have been shown to have high influence on the synchronizability of 
the network \cite{Motter-spectra}. Similarly, multiplicities of eigenvalues, 
particularly at 0 and 1, have direct relations with the properties of graphs 
\cite{lap-spectra,Jost}.

Following our recent works, where we used random matrix theory (RMT) to study 
spectral properties of 
adjacency matrix of various networks \cite{pap1}, in this paper we investigate 
spectral properties of Laplacians of networks under RMT framework. Particularly we 
study nearest neighbor spacing distribution (NNSD) of Laplacian matrix of various 
model networks, namely scale-free, small world and random networks. We find that 
inspite of spectral densities of different model networks are different, their 
eigenvalue fluctuations are same and follow Gaussian orthogonal ensemble (GOE) 
distribution of RMT. We attribute 
this universality to the presence of the {\it similar amount of randomness} in all 
these networks, and show that randomness in the network connections can be 
quantified by the Brody parameter coming from RMT.  
Furthermore, there exists one to one correlation between the diameter of the 
network and the eigenvalues fluctuations of the Laplacian matrix. By changing 
number of connections in the network we get transition to the GOE distribution. As 
Erd\"os and R\'enyi observed that with the fine tuning of network parameter all 
nodes get connected with a sudden transition; under the RMT framework our analysis 
suggests transition to some kind of spreading of randomness over the whole network.

Note that in this paper we consider normalized Laplacians, though for the RMT 
analysis form of Laplacians does not matter, because unfolding, a method to 
separate out system dependent part from the eigenvalues to study their universal 
behavior, removes scaling effects caused by the normalization factor. We consider 
normalized Laplacians because eigenvalues properties of this form of Laplacian are 
extensively investigated \cite{Lap-norm-Chung2,Lap-norm-others,Lap-norm-Internet,Lap-norm-CancerCell,Lap-norm-JHuang-GraphPartition}. 
They have been found to be an excellent candidate as a concise fingerprint of 
internet-like graphs \cite{Lap-norm-Internet} and graphs for the cancer cells 
\cite{Lap-norm-CancerCell}. Furthermore for some applications, such as graph 
partitioning, this normalized form is preferred over other forms of Laplacian 
\cite{Lap-norm-JHuang-GraphPartition, Jost}.

The paper is organized as follows: after this introductory section, in Sec. 
\ref{RMT}, we describe some techniques of RMT which we have used in our analysis. 
In Sec. \ref{analysis}, we analyze the NNSD of Laplacian for various networks, 
namely small-world, scale-free and random networks. In section \ref{transition} we 
study the effect of random connections on the level statistics. In this section we 
show the one to one correlation between GOE transition and small-world behavior. 
Finally, in Sec. \ref{summary}, we summarize and discuss about some possible future 
directions.

\section{Random matrix statistics} 
\label{RMT}

RMT was proposed by Wigner to explain the statistical properties of nuclear spectra 
\cite{mehta}. Later this theory had successfully been applied in the study of 
different complex systems such as disordered systems, quantum chaotic systems where 
RMT tells whether corresponding classical system is regular or chaotic or a mixture 
of both, spectra of large complex atoms, etc \cite{rev-rmt}. RMT is also shown to 
be of great interest in understanding the statistical structure of the empirical 
cross-correlation matrices appearing in the study of multivariate time series. The 
classical complex systems where RMT has been successfully applied are stock market 
(cross-correlation matrix is formed by using the time series of price fluctuations 
of different stock) \cite{rmt-stock};  brain (matrix is constructed by using EEG 
data at different locations) \cite{rmt-brain}; patterns of atmospheric variability 
(cross-correlations matrix is generated by using temporal variation of various 
atmospheric parameters) \cite{rmt-atmosphere}, etc.

We study eigenvalues fluctuations of the Laplacian of the networks. The eigenvalues 
fluctuations are generally obtained from the nearest neighbor spacing distribution 
(NNSD) of the eigenvalues. The NNSD follows two universal properties depending upon 
the underlying correlations among the eigenvalues. For correlated eigenvalues, the 
NNSD follows Wigner-Dyson formula of Gaussian orthogonal ensemble (GOE) statistics 
of RMT; whereas, the NNSD follows Poisson statistics of RMT for uncorrelated 
eigenvalues.

Here we briefly describe some aspects of RMT which we have used in our network 
analysis. We denote the eigenvalues of network Laplacian by 
$\lambda_i,\,\,i=1,\dots,N$, where $N$ is the size of the network and 
$\lambda_{i+1} > \lambda_i, \, \, \forall i$. In order to get universal properties 
of the fluctuations of the eigenvalues, it is customary in RMT to unfold the 
eigenvalues by a transformation $\overline{\lambda}_i = \overline{N} (\lambda_i)$, 
where $\overline{N}$ is the averaged integrated eigenvalue density \cite{mehta}. 
Since we do not have any analytical form for $\overline{N}$, we numerically unfold 
the spectrum by polynomial curve fitting. After the unfolding, the average spacing 
will be {\it unity}, independent of the system. Using the unfolded spectra, we 
calculate the spacing as $s_i=\overline{\lambda}_{i+1}-\overline{\lambda}_i$. The 
NNSD $P(s)$ is defined as the probability distribution of these $s_i$'s. In case of 
Poisson statistics, $P(s)=\exp(-s)$; whereas for GOE
\begin{equation}
P(s)=\frac{\pi}{2}s\exp \left(-\frac{\pi s^2}{4}\right).
\label{goe}
\end{equation} 
For the 
intermediate cases, the spacing distribution is described by Brody 
distribution \cite{brody}:
\begin{equation}
P_{\beta}(s) = A s^\beta\exp\left(-\alpha s^{\beta+1}\right),
\label{eq-brody}
\end{equation}
where
\begin{equation}
A = (1+\beta)\alpha~\mbox{and}~ \alpha = 
\left[\Gamma \left( \frac{\beta+2}{\beta+1}\right)\right]^{\beta+1}.
\end{equation}
This is a semiempirical formula characterized by the parameter $\beta$. As $\beta$ 
goes from $0$ to $1$, the Brody distribution smoothly changes from Poisson to GOE. 
We fit the spacing distributions of different networks by Brody distribution 
$P_{\beta}(s)$. This fitting gives an estimation of $\beta$, and consequently 
identifies whether the spacing distribution of a given network is Poisson or GOE or 
intermediate of these {\it two}.

\section{Laplacian matrix spectrum of complex networks}
\label{analysis}

Following we present the ensemble averaged spectral density and spacing 
distribution of random, scale-free and small-world networks.

\subsection{Random network}

\begin{figure}[t]
\centerline{
\includegraphics[height=6cm,width=7cm]{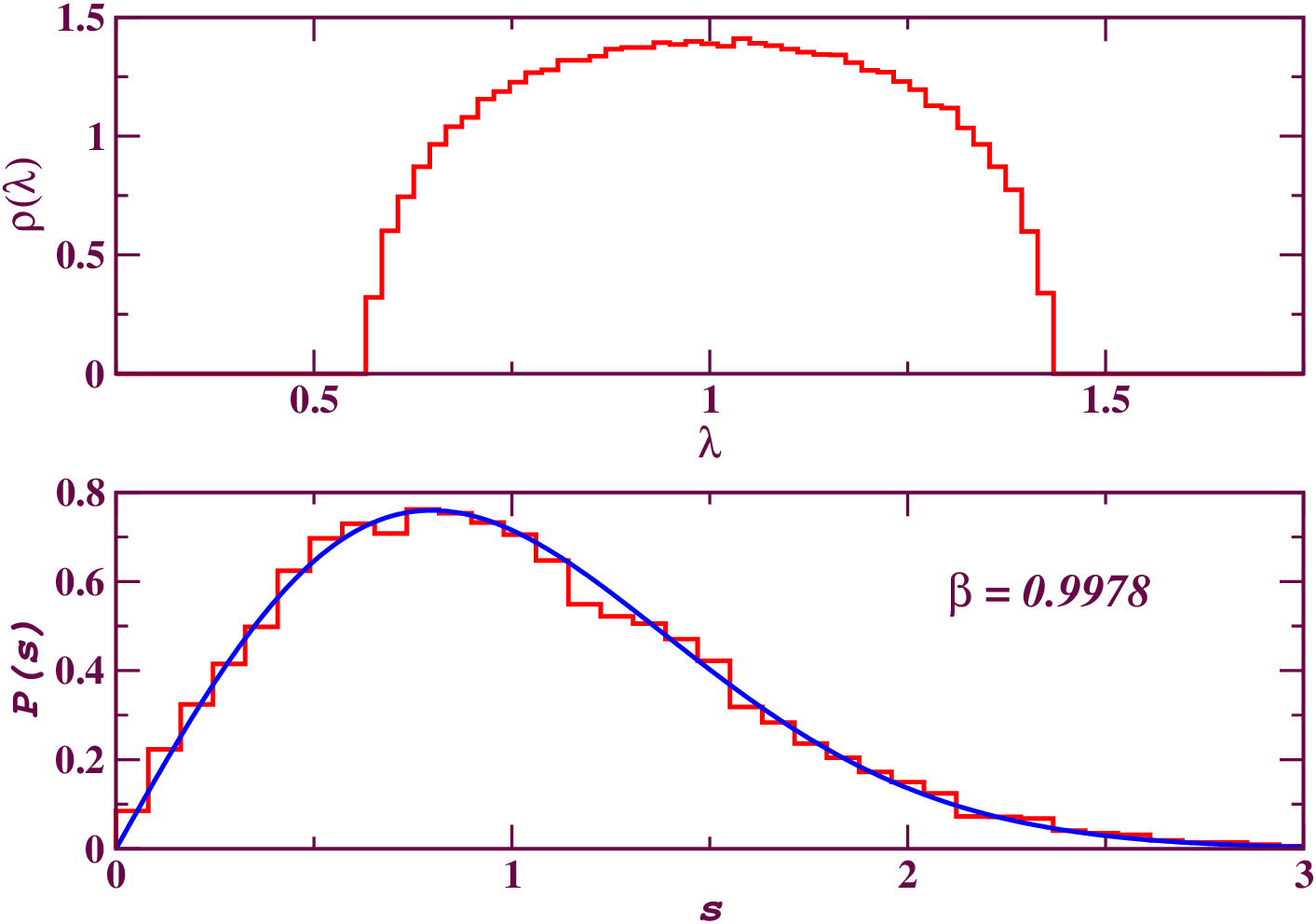}}
\caption{(Color online) (a) and (b) show the spectral density 
($\rho(\lambda)$) and corresponding spacing distribution ($P(s)$) for 
random network. The histograms are numerical results and the solid lines 
represent fitted Brody distribution. Figures are plotted for average over 
10 random realizations of the networks. All networks have $N=2000$ nodes 
and an average degree $k = 20$ per node.}
\label{rand-eig-L}
\end{figure}

First we consider random network generated by using Erd\"os and R\'enyi algorithm. 
We take $N=2000$ nodes and make random connections between pairs of nodes with 
probability $p=0.01$. This method yields a connected network with average degree $p 
\times N = 20$. Note that for very small value of $p$ one gets several unconnected 
components. Here, the choice of $p$ is such that it should be high enough ($p > 
p_c$) to give large connected component typically spanning all nodes \cite{note2}; 
and since most of real world networks are very sparse \cite{rev-network}, $p$ is 
small enough to have a sparse network. Laplacian is constructed using 
Eq.~(\ref{lap}). Because of the normalized form, the eigenvalues would always be 
within $0$ and $2$. Figure~\ref{rand-eig-L}(a) plots the spectral density of the 
Laplacian. Note that the distribution is averaged over 10 realizations of the 
network. The spectral density follows Wigner-Dyson semicircular distribution 
\cite{Lap-norm-Chung-RandomGraph}. To get the spacing behaviors, first the 
eigenvalues are unfolded by using the technique described in Sec. \ref{RMT}. This 
method yields the eigenvalues with constant spectral density on the average. These 
unfolded eigenvalues are used to calculate NNSD. The same procedure is repeated for 
an ensemble of the networks generated for different random realizations. 
Figure~\ref{rand-eig-L}(b) plots ensemble average of NNSD. By fitting this ensemble 
averaged NNSD with the Brody formula given in Eq.~(\ref{eq-brody}) we get an 
estimation of the Brody parameter $\beta = 0.9786 \sim 1$. This value of Brody 
parameter clearly indicates the GOE behavior of the NNSD [Eq.~(\ref{goe})].

\subsection{Scale-free network}

\begin{figure}[t]
\centerline{
\includegraphics[height=6cm,width=7cm]{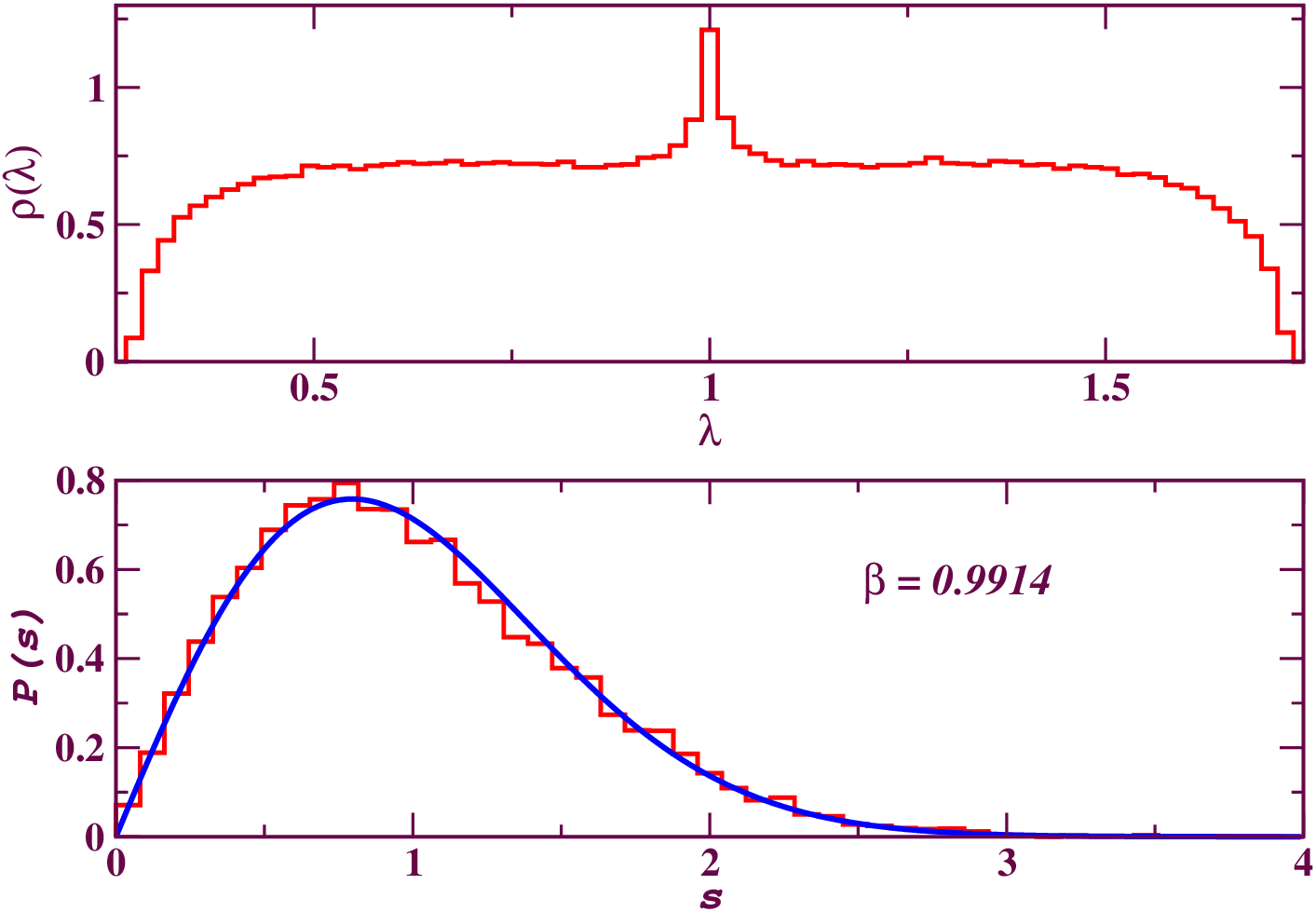}
}
\caption{(Color online) (a) and (b) show the spectral density ($\rho(\lambda)$) and 
corresponding spacing distribution ($P(s)$) of scale-free network. The histograms 
are numerical results and the solid lines represent fitted Brody distribution. 
Figures are plotted for average over 10 random realizations of the networks. All 
networks have $N=2000$ nodes and an average degree $k = 20$ per node.}
\label{SF-eig-L}
\end{figure}

Figure~\ref{SF-eig-L} plots the spectral density and the nearest neighbor spacing 
distribution of the Laplacian of scale-free network. Scale-free network is 
generated by using the model of Barab\'asi {\it et al.} \cite{BA}. Starting with a 
small number, $m_0$ of the nodes, a new node with $m \leq m_0$ connections is added 
at each time step. This new node connects with node $i$ with the probability 
$\pi(k_i) \propto k_i$ (preferential attachment), where $k_i$ is the degree of the 
node $i$. After $\tau$ time steps the model leads to a network with $N=\tau + m_0$ 
nodes and $m \tau$ connections. This model leads to a scale-free network, i.e., the 
probability $P(k)$ that a node has degree $k$ decays as a power law, $P(k) \sim 
k^{-\lambda}$, where $\lambda$ is a constant and for the type of probability law 
$\pi(k)$ that we have used $\lambda = 3$. Other forms for the probability $\pi(k)$ 
are possible which give different values of $\lambda$ and we find results similar 
to the ones reported here. Density distribution of the network has a very uniform 
distribution for almost all eigenvalues except for few lowest and highest ones. 
There is a slight peak around $one$ which corresponds to the peak at $zero$ for the 
adjacency matrix (\cite{pap1}). To calculate NNSD, we follow same procedure as 
described in the previous section. Fig.~ \ref{SF-eig-L}(b) shows that spacing 
distribution of this scale-free network follow GOE very closely $(\beta = 0.9914 
\simeq 1)$.

\subsection{Small-world network}

Figure~\ref{SW-eig-L} is plotted for small-world network. Small-world networks are 
constructed using the following algorithm by Watts and Strogatz \cite{SW}. Starting 
with a one-dimension ring lattice of $N$ nodes in which every node is connected to 
its $k$ nearest neighbors, we randomly rewire each connection of the lattice with 
the probability $p$ such that self-loop and multiple connections are excluded. Thus 
$p=0$ gives a regular network and $p=1$ gives a random network. The typical small 
world behavior is observed around $p=0.005$. We take $N=2000$ and average degree 
$k=40$. Spectral density of this network is complicated with several peaks. One 
peak is at $\lambda = 1$ which corresponds to the $\rho(0)$ peak for the spectra of 
adjacency matrix. For different values of $k$ the exact positions of other peaks 
may vary but overall form of spectral density remains similar. 
Fig.~\ref{SW-eig-L}{b) shows that spacing distribution of small world network 
follows GOE very closely $(\beta = 0.9584 \simeq 1)$.

\begin{figure}[t]
\centerline{
\includegraphics[height=6cm,width=7cm]{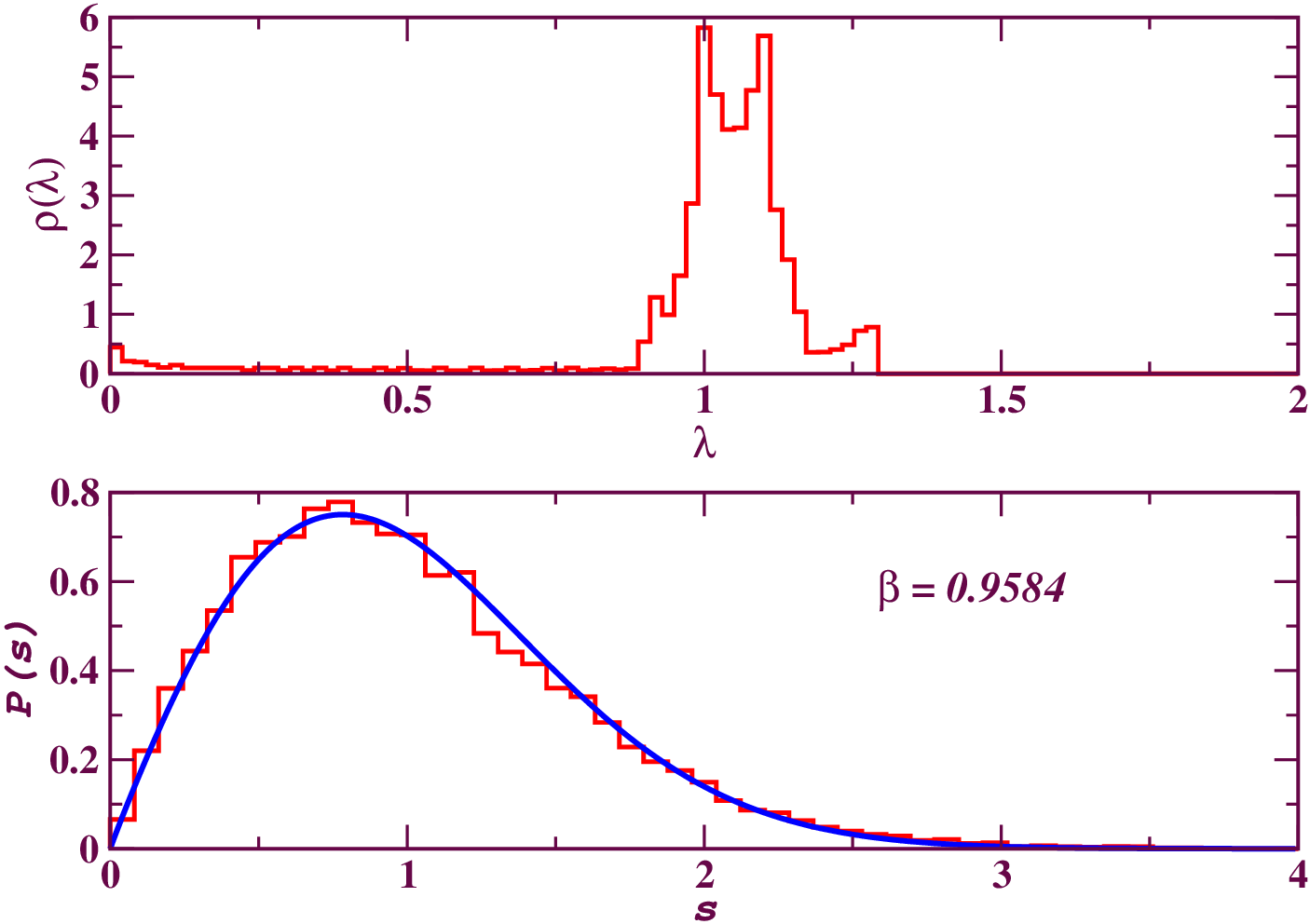}
}
\caption{(Color online) (a) and (b) show the spectral density ($\rho(\lambda)$) and 
corresponding spacing distribution ($P(s)$) of small-world network. The histograms 
are numerical results and the solid lines represent fitted Brody distribution. 
Figures are plotted for average over 10 random realizations of the networks. All 
networks have $N=2000$ nodes and an average degree $k = 20$ per node.}
\label{SW-eig-L}
\end{figure}

Figs.~\ref{rand-eig-L}(b), \ref{SF-eig-L}(b) and \ref{SW-eig-L}{b) show that 
spacing distribution of all the three networks follow GOE very closely $(\beta 
\simeq 1)$. Since spectral density of random network is very close to the 
semicircular, and in RMT literature semicircular distributions are extensively 
studied and are shown to give GOE statistics of the spacings, we expected that NNSD 
of random networks would follow $GOE$ as well. But for scale-free network and 
small-world network, density distributions have very different forms, still NNSD of 
these networks follow GOE statistics. This is a very interesting result. Following 
RMT these results imply that even though the spectral density of the scale-free and 
small-world network differ than the spectral density of the random network, but the 
correlations among the eigenvalues are strong enough to yield GOE statistics.

\begin{figure}[t]
\centerline{
\includegraphics[height=6cm,width=7cm]{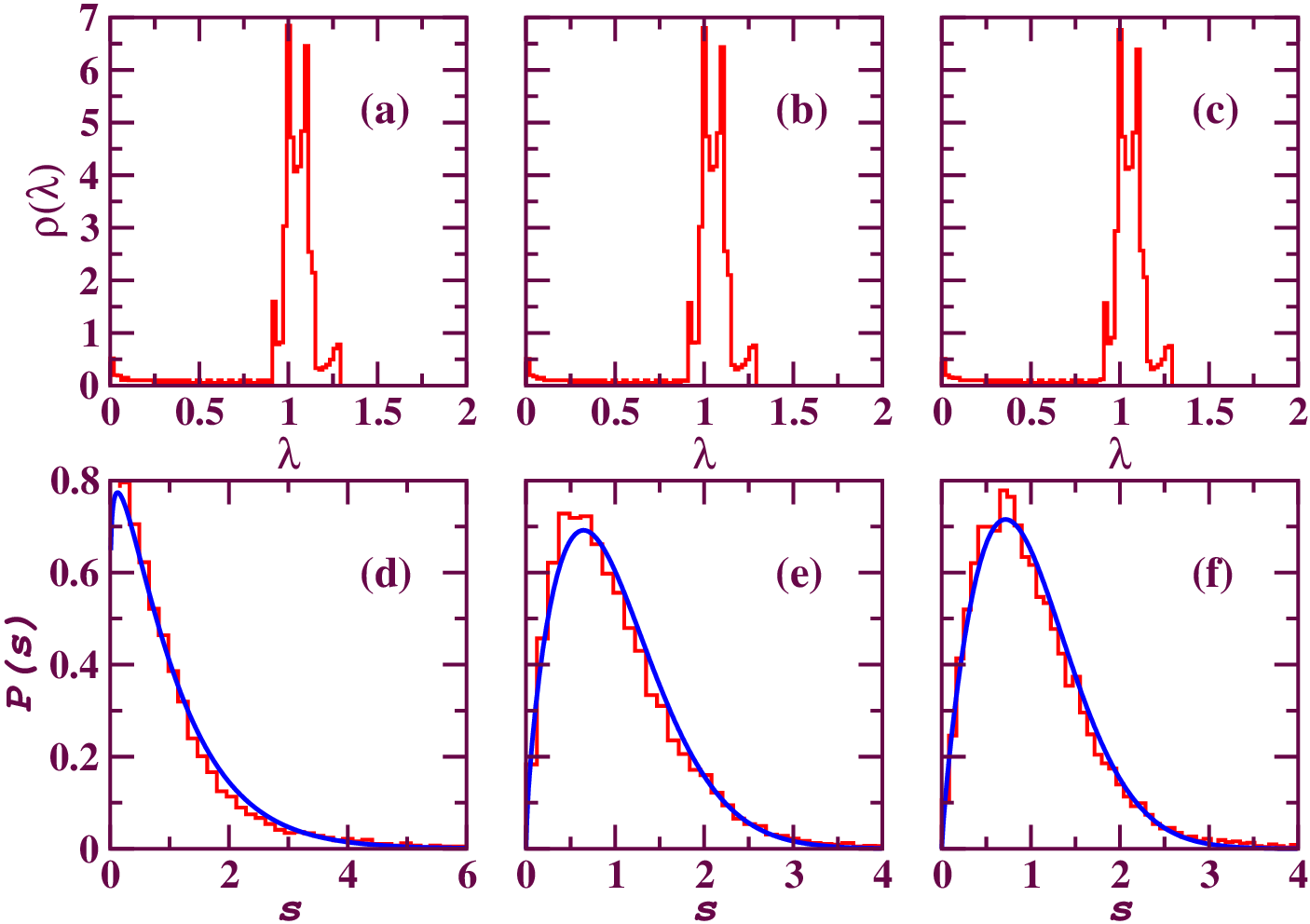}
}

\caption{Figures shows the transition from Poisson to the GOE statistics of the 
eigenvalues of the Laplacian of the network as a function of random rewiring 
parameter $p$. We take regular lattice of size $N=2000$ and average degree $k=40$. 
Links are rewires with probability $p$. Figures are plotted for average over 10 
random realization of the networks. } \label{regular-SW} 
\end{figure}

\section{Transition to GOE statistics and small-world behavior}
\label{transition}

All the above networks have some amount of random connections in them. In this 
section we discuss the networks which are regular and analyze the spectral 
properties as we go from regular network to the random one. Starting with a ring 
lattice of nodes $N = 2000$ and average degree $k=40$, we then randomize some 
connections with probability $p$, and observe the change in the spectral density 
and NNSD of the network as a function of $p$. For the original ring lattice, the 
underlying adjacency matrix would be a band matrix with entries one in the band, 
except diagonal elements which would be zero. Corresponding Laplacian would be a 
band matrix as well, with off-diagonal elements $-1/\sqrt{k \times k}$ 
(from Eq.~(\ref{lap})), where $k$ being the degree of each node, and diagonal elements
$1$. As a result of rewiring, few randomly 
selected pairs of nodes get connected, yielding few nonzero entries outside to the band 
of the corresponding Laplacian, while making few band elements zero.
For a completely random network, the corresponding Laplacian would 
have non zero random entries. 

Following we discuss the spectral behavior of Laplacian matrix as 
number of random connections in the underlying network is increased. 
Figures~\ref{regular-SW} (a) and (d) plot density distribution 
and NNSD for the $p = 5\times 10^{-5}$ case. At this value of $p$ there is as few 
as {\it one} random rewiring in the regular lattice. These figures shows that 
spectral density of the lattice is complicated without having any known form; and 
its spacing distribution closely follows Poisson statistics $(\beta \sim 0.1)$. We 
then increase the value of $p$ and randomize a fraction $p = 2 \times 10^{-4}$ of 
the edges. For this value of $p$, the spectral density and the spacing distribution 
are plotted respectively in Figure~\ref{regular-SW} (b) and (e). These figures 
reveal that for this small value of $p$, density distribution of the network does 
not show any noticeable change, whereas spacing distribution shows different 
property (corresponding to the $\beta \sim 0.67 $). As $p$ is further increased to 
$5 \times 10^{-4}$ the spectral density hardly shows any change but interestingly 
spacing distribution shows significantly different property than the Poisson 
statistics. Now the spacing distribution looks more like the intermediate between 
Poisson and GOE. By fitting the spacing distribution corresponding to this $p$ 
value with the Brody formula, we estimate the Brody parameter $\beta$ as $ \beta 
\sim 0.8$. As marked change in the Brody parameter for very small change in the 
random connections in the network (corresponding to $p = 5 \times 10^{-5}$, 
Fig.~\ref{regular-SW} (d)), we had expected that the Brody parameter will reach 
asymptotically to unity as we increase $p$. But the surprising finding is that the 
onset of transition from Poisson to GOE occurs at very small value of $p$ (Figs. 
~\ref{regular-SW} (e) and (f)), and is related with the SW properties of the 
networks. We calculate the network diameters and clustering coefficients for these 
values of $p$, which are listed below:

\begin{center}
\begin{table}[h]
\caption{Diameter, clustering coefficient, and Brody parameter
for the various values of rewiring probability $p$.}

\centerline{\begin{tabular}{|l|cc|cc|cc|cc|}
\hline
& $p$ && $L(p)/L(0)$ && $C(p)/C(0)$ && $1-\beta$&\\
\hline 
(a) & $5 \times 10^{-5}$ && $0.83925$ && $0.99980$ && 0.89463 & \\
(b) & $2 \times 10^{-4}$ && $0.53680$ && $0.99925$ && 0.33368 & \\
(c) & $5 \times 10^{-4}$ && $0.39776$ && $0.99857$ && 0.20131 & \\
\hline
\end{tabular}}
\label{brody-SW}
\end{table}
\end{center}
Table \ref{brody-SW} shows that for (b) and (c), the average length of the network 
is as small as the corresponding random network ($L_{random} \sim \ln(N)/\ln(k)$, 
and clustering coefficient is as high as the regular lattice ($ C_{regular} \sim 
3/4$), which are the properties of small-world networks. Values of $\beta$ reveals 
that Poisson to GOE ($\beta \sim 0 \rightarrow \beta \sim 1$) transition and small-world 
transition takes place for the similar value of $p$.

\section{Conclusions}
\label{summary}

In summary, we study eigenvalues spacing distribution of Laplacian of the model 
networks studied extensively in the literature. They follow universal GOE 
statistics. From RMT analogy it tells that there exists correlations between the 
eigenvalues of the network arising from certain symmetries in the interactions. We 
attribute this universality to the {\it similar} amount of randomness in the model 
networks, which may arise naturally, in order to capture the real world network 
properties. We study the effect of the randomness in the network connections on the 
eigenvalues fluctuations of network Laplacian, and use Brody parameter to quantify 
this randomness. These studies reveal that there is a direct relation between the 
random connections and the Brody parameter. For the regular network (with the 
average degree greater then some value) we get Poisson distribution, as we make 
random rewiring of the connections the spacing distribution start deviating from 
the Poisson statistics, first it shows intermediate of Poisson and GOE statistics 
and for {\it sufficiently large} number of random connections it show almost GOE 
statistics ($\beta \sim 1$). According to the interpretation of RMT, at this value 
of $p$, eigenvalues are as much correlated as for the completely random networks. 
Furthermore we find that this transition to GOE statistics happens at the onset of 
the small world behavior.

Universal RMT results shown by the networks suggest that they can be modeled as a 
random matrix chosen from GOE of RMT. Eigenvalues fluctuations following GOE 
statistics argue that there exists some kind of spreading over the randomness in 
the whole network, which may be essential for robustness of the system. According 
to many recent studies, randomness in the connections is one of the most important 
and desirable ingredients for the proper functionality or the efficient performance 
of the system having underlying network structure. For instance, information 
processing in the brain is considered to be highly influenced by random connections 
among different modular structure \cite{face}. Our analysis suggests that {\it 
randomness} in the complex networks can be studied under the RMT framework. 
Furthermore, Laplacian spectra have been investigated to understand various 
dynamical processes on networks \cite{syn-Atay,syn-spectra,Motter-spectra} as well, 
hence the RMT analysis of network Laplacians could also be important in view of 
these demonstrations of the relations between spectral properties and dynamical 
processes.

Following RMT analysis of adjacency matrices of networks introduced in our previous 
work \cite{pap1}, we investigate random matrix properties of network Laplacians. In 
this paper we only concentrate on the spacing distribution of the eigenvalues of 
the Laplacian, which explains short range correlations among the eigenvalues. 
Further investigations would involve more sensitive analysis like $\Delta_3$ 
statistics to understand the long range correlations \cite{pap4}.

\section{Acknowledgments} SJ acknowledges Prof. J\"{u}rgen Jost for discussing the 
importance of normalized Laplacians and a very enlighting 2006 summer course in 
Max-Planck-Institut f\"{u}r Mathematik in den Naturwissenschaften, Leipzig.


\begin{thebibliography}{99}
\bibitem{note}Relation between the spectra of $L$ and $A$ is known only in the case 
of $d$-regular (degree of each node of the network being equal, $d_i=d \, \, \forall i$) 
graphs. In this very special case, $L = I - A/d$, where $I$ is a unit matrix, and 
hence $\lambda_L = 1 - \lambda_A/d$, where $\lambda_L$ and $\lambda_A$ are the 
eigenvalues spectra of $L$ and $A$ respectively. For this case, $L$ and $A$ are 
linearly related, and therefore the statistical propertes of their spectra would be 
same. If the degree of nodes are not same, as usually happens in complex networks, 
then relation between the eigenvalues of $L$ and $A$ is not known.
\bibitem{note2}Before this critical value of $p \le p_c$, one gets several
disconnected components. Laplacian matrix $L$ of a network having several
disconnected components can be written as Kronecker or direct sum of the Laplacian
matrices $L_i$ of the disconnected components, i.e., $L = \bigoplus_{i=1}^{N}\,L_i$, 
where
$N$ is the number of disconnected components. In this case, even if the NNSD of the
individual components follows GOE, but in general the NNSD of $L$ does not follow GOE
property.
\end{thebibliography}
\end{document}